\documentclass[a4paper,11pt]{article}

\usepackage{amsmath}
\usepackage{amssymb}
\usepackage{epsfig}
\usepackage{color}

\newcommand{\fig}[1]{fig.~\ref{fig:#1}}

\newcommand{\figs}[1]{figs.~\ref{fig:#1}}

\newcommand{\eq}[1]{eq.~(\ref{eq:#1})}

\newcommand{\eqs}[1]{eqs.~(\ref{eq:#1})}

\newcommand{\TeV}{\,\mathrm{TeV}}

\newcommand{\MeV}{\,\mathrm{MeV}}
\newcommand{\keV}{\,\mathrm{keV}}
\newcommand{\eV}{\,\mathrm{eV}}

\newcommand{\capdef}{}
\newcommand{\mycaption}[2][\capdef]{\renewcommand{\capdef}{#2}%
        \caption[#1]{{\itshape #2}}}
\makeatletter
\renewcommand{\fnum@table}{\textbf{\tablename~\thetable}}
\renewcommand{\fnum@figure}{\textbf{\figurename~\thefigure}}
\makeatother

\newlength{\myem}
\newcommand{\sep}[1]{#1}
\newcounter{mysubequation}[equation]
\renewcommand{\themysubequation}{\alph{mysubequation}}
\newcommand{\mytag}{\stepcounter{mysubequation}%
\tag{\theequation\protect\sep{\themysubequation}}}
\newcommand{\globallabel}[1]{\refstepcounter{equation}\label{#1}}

\newcommand{\Ye}{Y_{L_e}}
\newcommand{\YLe}{Y_{L_e}}

\newcommand{\YLt}{Y_{L_\tau}}
\newcommand{\Ynut}{Y_{\nu_\tau}}

\newcommand{\Vt}{V_{\tau}}
\newcommand{\mmR}{m^2\! R}
\newcommand{\nue}{\nu_e}
\newcommand{\nueb}{\bar\nu_e}
\newcommand{\num}{\nu_\mu}
\newcommand{\numb}{\bar\nu_\mu}
\newcommand{\nut}{\nu_\tau}
\newcommand{\nutb}{\bar\nu_\tau}
\newcommand{\nB}{n_{B}}
\newcommand{\mun}{\mu_{\nu_e}}
\newcommand{\mut}{\mu_{\nu_\tau}}

\newcommand{\GF}{G_{\text{F}}}
\newcommand{\SN}{SN 1987A}
\newcommand{\km}{\,\mathrm{km}}
\renewcommand{\sec}{\,\mathrm{sec}}

\newcommand{\Msun}{M_{\odot}}

\newcommand{\meff}{m_{\text{eff}}}

\newcommand{\Plossbar}{P(\bar\nu_\tau\to\text{bulk})}

\newcommand{\sud}{s_{12}}
\newcommand{\cud}{c_{12}}
\newcommand{\sdt}{s_{23}}
\newcommand{\cdt}{c_{23}}

\newcommand{\SNS}{Scuola Normale Superiore and INFN Sezione di Pisa \\
Piazza dei Cavalieri 7, I--56126 Pisa, Italy}

\newcommand{\CERN}{Theory Division, CERN, CH--1211 Geneva, Switzerland}

\newcommand{\preprintdate}{}
\newcommand{\preprintnumber}{%
CERN--TH/2003--038}
\newcommand{\titletext}{Signatures of Supernova Neutrino
  Oscillations\\ into Extra Dimensions}
\newcommand{\authortext}{\large G. Cacciapaglia$^{\, a}$, M. Cirelli$^{\, a}$,
  A. Romanino$^{\, b}$
  \medskip\\\em\normalsize $\mbox{}^a$ \SNS
  \\[0.1\baselineskip]
  $\mbox{}^b$ \CERN}

\newcommand{\abstracttext}{We consider the mixing of muon and tau
  neutrinos with sterile fermion fields propagating in extra
  dimensions in the context of core collapse supernova physics,
  extending the analysis of the electron neutrino case done in a
  previous work.  We show that the potentially dramatic modifications
  to the supernova evolution are prevented by a mechanism of feedback,
  so that no severe bounds on the parameters of the extra dimensions
  need to be imposed.  Nevertheless, the supernova core evolution is
  significantly modified.  We discuss the consequences on the delayed
  explosion mechanism and the compatibility with the \SN\ signal.
  Then, for the cases of both $\nu_{\mu,\tau}$ and $\nu_e$ mixing with
  bulk fermions, we analyse the distinctive features of the signal on
  Earth.}

\title{
\normalsize
 \begin{tabular}[t]{l}
 \preprintdate\end{tabular}
 \hspace*{\fill}
 \begin{tabular}[t]{l}\preprintnumber\end{tabular}
 \vspace{3\baselineskip}\\
\LARGE\bfseries\titletext\bigskip}
\author{\begin{minipage}[t]{0.8\textwidth}
\normalsize\centering\authortext
\end{minipage}}
\date{}

\begin{document}

\bigskip
\maketitle
\begin{abstract}\normalsize\noindent
\abstracttext
\end{abstract}\normalsize\vspace{\baselineskip}

\noindent

\section{Introduction}

The study of sterile neutrinos propagating in the bulk of extra space
dimensions~\cite{NUinED} and of core collapse supernova
evolution~\cite{reviews} establishes a nice link that can have
interesting implications on both sides, even though the role of
sterile neutrinos is not believed to be primary in solar and
atmospheric contexts~\cite{sno, SK}.  We consider indeed a
framework~\cite{LRRR} in which the solar and atmospheric oscillation
signals are independently accounted for in the standard way.  In
addition, a small mixing with the 4-dimensional Kaluza--Klein (KK)
states (the 4d correspondent of the extra dimensional field) is the
source of the non-standard effects on supernova (SN) physics, since it
implies resonant oscillations with ordinary Standard Model neutrinos
via matter effects.  The point is that we have a large number of
resonances that can enhance the effect, even if the mixing angles are
small.  As KK modes do not interact with ordinary matter and are
therefore not trapped, conversions lead to a faster cooling and
deleptonization of the SN dense core.  The observed \SN\
signal~\cite{signal} requires such an energy loss not to take place
with a time scale much shorter than $10 \sec$. However, the consequent
potentially severe bounds on the extra dimension size~\cite{BCS} are
widely relaxed by a feedback mechanism~\cite{LRRR2,CCLR} that
self-limits the conversion. In a previous work~\cite{CCLR}, we
analysed the possibility of a mixing between the electron neutrinos
and a sterile bulk fermion and we explicitly showed that the mechanism
indeed prevents the potentially dramatic cooling in a wide portion of
the parameter space. We also pointed out that new interesting effects
on the SN dynamics can arise in the enlarged parameter space.  We
speculated that the enrichment of the $\nue$ component in the outgoing
neutrino flux might help the delayed shock mechanism and drive the
explosion.\footnote{The difficulty to obtain an explosion in
  simulations, however, can be explained by the complexity of the
  system to be simulated, so that no new physics is in principle
  necessary.}  Furthermore, we found that while this picture is in
good agreement with the observed \SN\ signal, the flux composition can
be significantly modified.

\medskip

In the first part of the present work we address the case of muon and
tau neutrinos.  In the standard picture of collapse and cooling of the
SN core, their role is rather different from the electron ones.  In
fact, while electron neutrinos are copiously produced by the
neutronization process during collapse and build up a degenerate
$\nue$ sea with large chemical potential $\mun$ (therefore electron
antineutrinos are a negligible fraction in these conditions), muon and
tau neutrinos and antineutrinos are produced in pairs and have almost
vanishing chemical potential.  Moreover, lacking in charged current
interactions, they have a faster diffusion and give an important
contribution to the transport of thermal energy out of the core.
Also, their lepton number does not play a significant role in the
diffusion dynamics.

The mixing of tau or muon neutrinos with bulk fermions opens an
unconventional escape channel, which can affect the SN core cooling
and deleptonization in a peculiar way.  We will see that the usual
energy loss bounds can be relaxed, as in the electron case, by the
action of a feedback.  We will then be interested in the
phenomenological consequences on the revival of the shock and in the
implications for the \SN\ signal.

In the last part of the paper we discuss the peculiar signatures of
the considered effects on the neutrino signal on the Earth. This we do
not only for the case of $\num$, $\nut$ but also for the case of
$\nue$ conversion considered in~\cite{CCLR}.

\medskip

In the following we will assume that the sterile fermions mix
predominantly with the tau neutrino. This is the case if the SM/bulk
neutrino couplings reflect the hierarchical structure of the SM Yukawa
couplings.  Actually, the inclusion of a significant muon neutrino
mixing with bulk fields would just unnecessarily complicate the core
dynamics. The point is that the oscillations into bulk neutrinos build
up a non-zero tau (or muon) lepton number. The tau neutrino
abundance is simply determined by the tau lepton number, since tau
leptons are not thermally produced, as their mass is large with
respect to the energies involved ($T \sim 30 \MeV$).  On the contrary,
the muon neutrino abundance would also depend on the non-negligible
muon fraction through the $\beta$-equilibrium processes $\mu^-\, p
\leftrightarrow n\, \num$ and $\mu^+\, n \leftrightarrow p\, \numb$.
One can also wonder whether the $\num$--$\nut$ mixing could
reintroduce the muon fraction in the problem.  This is not the case,
since that mixing is largely suppressed by the difference at one loop
of the muon and tau matter potential~\cite{mutau} and by an even
larger difference originating from the new effect.  In fact, as soon
as the resonant conversion into the bulk begins, a non-vanishing tau
lepton number is generated and thus the matter interactions produce
different effective masses for tau and muon (anti)neutrinos.  For
sufficiently large mixing with the extra dimension fields, as is our
case, the oscillations are suppressed before they significantly
influence the neutrino fractions.  Notwithstanding the complications
described above, we do not expect a qualitatively different
behaviour in the presence of muons; in section \ref{sec:phenomenology}
we will thus deduce the case of $\num$ mixing with bulk fermions
from our results.

\medskip

The paper is organized as follows.  In section \ref{sec:framework} we
briefly describe the general extra dimensional scenario and set the
parameter space in which we are interested.  In section
\ref{sec:model} we first sketch the expected effect of such new
physics on the evolution and then present the numerical model that we
use.  Finally, in section \ref{sec:phenomenology} we address the new
phenomenology based on the results of our calculation and we compare
the expected neutrino flux on Earth in the cases of $\nue$, $\num$ and
$\nut$ conversions.  In section \ref{sec:conclusions} we sum up.

\section{Framework}
\label{sec:framework}

The cornerstones of our large extra dimension scenario are the
following.  We consider a single gravitational extra
dimension~\footnote{More extra dimensions could be present at higher
  energies. This scenario is compatible both with a fundamental
  gravity scale at $\sim 1~\TeV$~\cite{TeV} and with astrophysical and
  cosmological bounds~\cite{bounds}.}, probed by a sterile fermionic
field, which results from a 4d point of view as a KK tower of states,
predominantly mixed with tau neutrinos.

\begin{figure}[!t]
\begin{center}
\epsfig{file=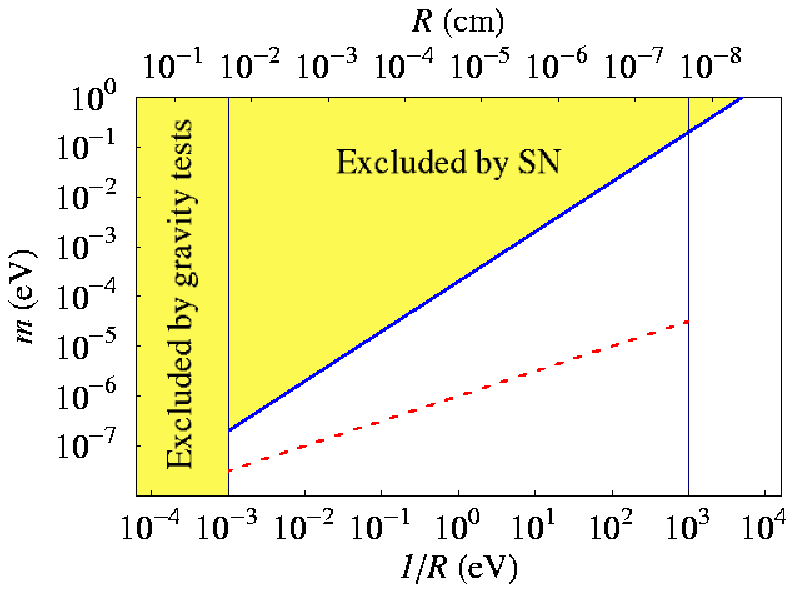,width=0.6\textwidth}
\end{center}
\mycaption{Excluded regions in the $m$--$R$ plane, for the case of
  $\nut$ ($\num$) escape into the bulk. The red dashed line
  (corresponding to $\mmR = 10^{-12}\, \eV$) marks the lower border of
  the region where the supernova core evolution is sensibly modified.
  In the region on the right ($1/R > 1 \keV$) the effect is not
  active: neutrinos do not cross a sufficient number of
    resonances.}
\label{fig:bounds1}
\end{figure}

The parameter space we are interested in is illustrated in
\fig{bounds1}.
The radius $R$ of the extra dimension, which sets the scale of the KK
spectrum, is allowed to lie in the broad range
$10^{-3}\eV\lesssim\frac{1}{R}\lesssim 1\keV$.  The lower bound comes
from direct gravity probes, the upper one from requiring that a
sufficient number of resonances be crossed by neutrinos travelling in
the core, a condition for observable effects to arise.  The mixing
angle of the $\nut$ with the {\it k}-th eigenstate of mass $M_k$ is
generically parametrized as $\theta_k \simeq \frac{m}{\sqrt{2} M_k}$,
which defines the parameter $m$; we only assume that the density of KK
states is proportional to $R$, a feature that is quite
model-independent. The condition $mR \ll 1$ ensures the smallness of
the mixings, while keeping subleading effects under
control~\cite{CCLR} requires $mR \ll 2 \times 10^{-4}$.  The above
requests identify the unshaded region in \fig{bounds1}. In this region
the new effects enter the analysis only through the crucial parameter
$m^2R$.

If feedback effects were not taken into account, only values of $\mmR$
up to $10^{-10}\eV$ would be compatible with the energy loss argument.
For example, for $\mmR=5\times 10^{-5}\eV$, the core would cool down
in about $10^{-5}\sec$. On the contrary, including the feedback, any
value of $\mmR$ up to $5 \times 10^{-5}\eV$ turns out to be
acceptable, while beyond that point the core is cooled by otherwise
subleading effects.  Thus, a new portion of the parameter space,
$10^{-10}\eV\lesssim \mmR\lesssim 5 \times 10^{-5}\eV$, is opened once
the feedback is taken into account.~\footnote{Compared with the
  electron neutrino escape case, the overall shift to higher values is
  justified (i) by the smaller number density of $\nutb$ with respect
  to $\nue$, since they are thermal, not degenerate, at the beginning;
  (ii) marginally, by the fact that $\nut$ diffusion is faster.}  In
the following we will focus on the range
\begin{equation}
  \label{eq:mmRrange2}
  10^{-12}\eV\lesssim \mmR\lesssim 5 \times 10^{-5}\eV\, ,
\end{equation}
where the neutrino escape affects the core evolution in a sizeable
way.

\medskip

Tau neutrinos in the SN core experience a
Mikheyev--Smirnov--Wolfenstein (MSW) potential given by
\begin{equation}
  \label{eq:V}
  \Vt =
  \sqrt{2}\GF\nB\left(\frac{1}{2}Y_e+2Y_{\nut}+Y_{\nue}-\frac{1}{2}\right) \, ,
\end{equation}
where $\GF$ is the Fermi constant, $\nB$ is the baryon number density,
and $Y_x$ is the net number fraction per baryon of the species $x$:
$Y_x=(N_x - N_{\bar x})/\nB$.  It will also be useful to consider the
lepton number fractions
$Y_{L_{e,\mu,\tau}}=Y_{e,\mu,\tau}+Y_{\nue,\num,\nut}$.

At the beginning $Y_e$ is the dominant term and, since it is not
expected to be larger than $0.4$ in the inner core \cite{reviews},
$\Vt$ is clearly negative.  The tau antineutrinos acquire an effective
squared mass $\meff^2 = 2 E_{\bar\nu} V$, where
$E_{\bar\nu}$ is the antineutrino energy. The effective mass
changes slightly along the neutrino path: whenever $\meff$ equals the
mass of one of the KK sterile states, a resonance occurs and the tau
antineutrino has a certain probability of oscillating into a bulk
state and escaping from the core. As a result, the tau lepton number
increases and the potential becomes less negative. If the potential
were positive the resonances would be met by neutrinos and the
potential would get smaller.  Note also that $\Ynut$ is zero at the
beginning but a positive value will be generated by the $\nutb$ escape
itself.  This term, which has a different weight for the $\num$, is
responsible for the suppression of the vacuum $\num$--$\nut$
oscillations.

The disappearance probability of a $\nutb$ into the bulk when
travelling a distance $L$ is given by the usual Landau--Zener formula
\begin{equation}
  \Plossbar \simeq L\,\frac{\pi}{2\sqrt{2}}\mmR
  \left(\frac{|\Vt|}{E_{\bar\nu}}\right)^{1/2} \,.
  \label{eq:Ploss}
\end{equation}
As is apparent, the parameter $\mmR$ sets the overall magnitude of the
escape effect.

\section{Model and evolution}
\label{sec:model}

Having introduced all the ingredients of the problem, let us first
discuss the basic features of the expected evolution, compared when
necessary with the case of $\nue$ conversion addressed in~\cite{CCLR}.

As already mentioned, the matter MSW potential is initially negative
in the whole inner core so that only $\nutb$ cross resonances.  As
conversions push $\Vt$ towards zero and deleptonization pulls it down
to its negative minimum value, a non-trivial interplay between
diffusion and the new effect takes place.  Nevertheless, the main
features of the evolution can be easily understood. First of all,
since the new physics is in the $\nut$ sector, the $\YLe$ evolution is
only indirectly affected, mainly via the modifications to the
temperature.  As the $\nue$'s are almost degenerate, we expect their
diffusion to be only slightly changed and in particular the
deleptonization time scale to be unaffected.  On the other hand, the
$\nutb$ escape generates a positive $Y_{\nut}$, the balance
$\nut$--$\nutb$ is broken and a positive chemical potential $\mut$
arises.  This in turn inhibits the escape itself, both because $\Vt$
is lifted towards zero by the term $Y_{\nut}$ in \eq{V} and, more
important, because the $\nutb$ abundance is suppressed in the presence
of the chemical potential.  In particular, $\Vt$ has no chances to
switch to positive values and only $\nutb$ do convert during all the
evolution.  Now, how do we expect the $T$ evolution to be affected? In
a first fast phase, as long as the conversion is dominant, a portion
of the thermal energy is absorbed both by bulk neutrinos and by the
sea of increasingly degenerate $\nut$'s. During all the following
evolution, the thermal energy that the $\nutb$'s escape keeps on
transferring to tau neutrinos is carried outside by their diffusion.
Thus, we expect the $T$ evolution to be sped up with respect to the
standard case.  Moreover, we expect the $\nutb$ escape cooling channel
to be more effective than in~\cite{CCLR} as the $\nut$ diffusion time
scale is shorter than for $\nue$'s.  Finally, the tau lepton number
that arises in the inner core diffuses out into the mantle and may
significantly influence the outgoing flux composition.

\medskip

To study the above picture in some more detail, we use a simplified
model that includes all the relevant physics, although it is not
intended as a complete description of the SN core dynamics.  A more
detailed analysis would require a complete simulation of the core and
of the external layers, with the inclusion of the new effect.

We follow all the core evolution in terms of the temperature $T$, the
electron leptonic fraction $\YLe$ and the tau leptonic fraction $\YLt$
(that is equal to the neutrino fraction $\Ynut$). One can express all
the other relevant observables (chemical potentials, entropy, fluxes,
\dots) simply as functions of these quantities, using thermodynamic
equilibrium, beta equilibrium and charge neutrality.  As
in~\cite{CCLR}, we consider a typical SN core characterized by a mass
of $1.5 \,\Msun$ and a radius of $12.7\km$.  For the matter density,
we use a static profile $\rho(r)$, justified by the relative
hydrodynamical stability of the inner core in which we are interested.
Explicitly $\rho(r)=\rho_c/(1+(r/\bar{r})^3)$, with $\rho_c=7.5\times
10^{14}$ g/cm$^3$ and $4/3 ~\pi ~\bar{r}^3 ~\rho_c = 1.1\,\Msun$.  As
for the equation of state, we consider a core made only of $n$, $p$,
$e^\pm$, $\nu_{e,\mu,\tau}$, $\bar\nu_{e,\mu,\tau}$ and $\gamma$.  The
effective nucleon mass is expressed as $ m_N^* = \frac{m_N}{1+\beta_0
  ~\rho/\rho_0}$, where $m_N = ~939 \MeV$ is the vacuum value,
$\beta_0$ is chosen to be 0.5 and $\rho_0 = 3\times 10^{14}$ g/cm$^3$
is the reference nuclear density.  As for the mean free paths
$\lambda_{\nu_e}$ and $\lambda_{\nu_\mu,\nu_\tau}$, in general they
would exhibit a non-trivial dependence on the neutrino energy $E_\nu$
and the evolution variables ($\rho$, $T$, $\YLe$, $\YLt$), besides on
several other aspects related to the medium in which the neutrinos
diffuse.  However, the essential features can be grasped by assuming a
simple inverse quadratic dependence on the neutrino energy for all
species and incorporating an inverse dependence on matter density for
muon and tau neutrinos, while keeping constant with density the mean
free path of electron neutrinos.  We checked these simplified
assumptions versus the more complete modellings of~\cite{Reddy}.
Moreover, the above choices allow us to obtain an evolution whose main
features and timescales agree with the results of more sophisticated
analyses. The expressions we use are the following:\footnote{With
  respect to the electron neutrino case of~\cite{CCLR} we refined the
  value of $\lambda^0_{\nu_\mu,\nu_\tau}$ and we introduced the
  density dependence for muon and tau neutrinos.}
\begin{equation}
  \label{eq:mfp}
  \lambda_{\nu_e}(E_\nu)=\lambda_{\nu_e}^0 \frac{E_{\nu,0}^2}{E_\nu^2}\,,
  \qquad  \lambda_{\nu_\mu,\nu_\tau}(E_\nu,r)
  =\lambda_{\nu_\mu,\nu_\tau}^0 \frac{\rho_{c}}{\rho(r)} \frac{E_{\nu,0}^2}{E_\nu^2}\,,
\end{equation}
taking $\lambda_{\nu_e}^0 = 1.2$ cm and $\lambda^0_{\nu_\mu,\nu_\tau}
= 2.8$ cm at the reference energy $E_{\nu,0} = 260 \MeV$ and reference
density $\rho_c=7.5\times 10^{14}$ g/cm$^3$.  The initial profiles for $T$
and $\YLe$ are typical ones as taken in~\cite{CCLR}. Of course, the
profile for $\YLt$ is zero in all the core at the beginning.

Finally, with all the above prescriptions (see~\cite{CCLR} for a more
detailed discussion), the evolution equations for $\Ye$, $\YLt$ and
$T$ read \globallabel{eq:evolution}
\begin{align}
  \nB\frac{\partial\Ye}{\partial t} &= \vec\nabla\cdot(a_e~\vec\nabla\mun) \mytag\\
  \nB\frac{\partial\YLt}{\partial t} &= \vec\nabla\cdot(a_\tau~\vec\nabla\mut)
 + \frac{1}{4\sqrt{2}\pi} \mmR\sqrt{|V_\tau|}~T^{5/2}
  F_{3/2}\left(-\frac{\mut}{T}\right) \mytag \\
  \begin{split}
    \nB T\frac{\partial s}{\partial t} &= a_e\left(\vec\nabla\mun\right)^2+
    a_\tau\left(\vec\nabla \mut\right)^2+
    \vec\nabla\cdot\left((a_e+a_\mu+a_\tau)\frac{\pi^2}{6}\vec\nabla T^2\right) \\
    &\quad-\frac{1}{4\sqrt{2}\pi} \mmR\sqrt{|V_\tau|}~T^{7/2}\left(
      F_{5/2}\left(-\frac{\mut}{T}\right)+\frac{\mut}{T}
      F_{3/2}\left(-\frac{\mut}{T}\right)\right)\,,
  \end{split} \mytag
\end{align}
where $a_i = \lambda^0_{\nu_i} E_{\nu,0}^2/(6\pi^2)$, $i= e, \mu, \tau$; $s$ is the entropy per baryon and the Fermi integrals are defined as $F_q(y) = \int_0^\infty dx \frac{x^q}{e^{x-y}+1}\,$.

One can recognize the standard transport equation for electron neutrinos in eq.~(\ref{eq:evolution}a), while eq.~(\ref{eq:evolution}b) for the tau neutrinos shows the additional escape term that comes from the integration over neutrino energies of the probability in \eq{Ploss}, for $L$ equal to a mean free path.
The first line of the temperature eq. (\ref{eq:evolution}c) features the heating terms $a_j\left(\vec\nabla\mu_{\nu_{j}}\right)^2$ ($j=e,\tau$) associated to the degradation of the degeneracy energy of neutrinos reaching regions with lower chemical potential.
The novelty is the presence of such a term for tau neutrinos also, since they possess a non-vanishing chemical potential now.
After the standard temperature diffusion term $\vec\nabla\cdot(\vec\nabla T^2)$, last comes the contribution from tau antineutrinos escape, which is a cooling term: its first part is related to the energy flux into bulk when an antineutrino is lost, the second one is associated with the (negative) tau lepton number flux into the bulk and thus with the storing of energy in the tau neutrino degenerate sea.

\medskip

\begin{figure}
\begin{center}
\epsfig{file=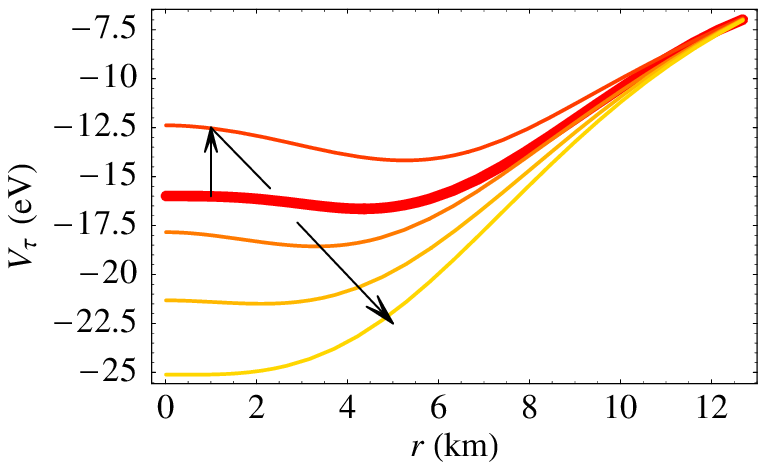,width=0.60\textwidth}
\end{center}
\mycaption{Profiles of the MSW potential for tau neutrinos. The thick
  line is the initial configuration corresponding to the typical
  initial profiles assumed in~\cite{CCLR}; the thin lines are
  snapshots at 1, 5, 10 and 20 secs, following the arrows, for the
  case $\mmR = 10^{-7}~eV$.}
\label{fig:VtauEv}
\end{figure}

The numerical solutions of the above equations, for several choices of
the parameter $\mmR$ in the range of \eq{mmRrange2}, confirm the
expectations presented at the beginning of this section.  Essentially,
the electron neutrino density follows the standard evolution while the
temperature fall is faster.  The tau leptonic fraction $\YLt$ rapidly
grows from zero up to a certain profile (the higher the larger $\mmR$,
but always below about 0.1) and then lowers on typical diffusion
timescales.  In parallel, the chemical potential $\mut$ quickly grows
up to values as large as those of $\mun$ (of order $200 \MeV$) and
then clears out with diffusion.  A typical evolution of the matter
potential is shown in \fig{VtauEv}: at the beginning it is pushed
towards zero by the rising of $\YLt$ but, the initial potential being
well below zero, the available energy is not sufficient for a complete
zeroing, and diffusion soon reverses the trend.

\section{Phenomenology}
\label{sec:phenomenology}

\begin{figure}
\begin{center}
\epsfig{file=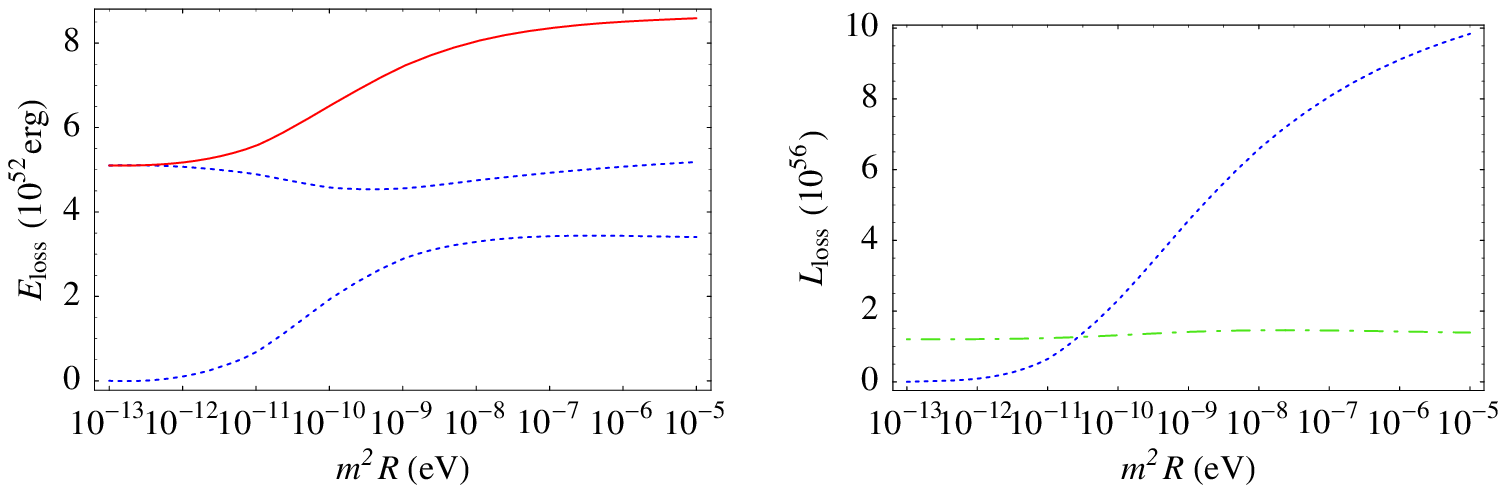,width=1.00\textwidth}
\end{center}
\begin{minipage}[t]{0.475\textwidth}
  \mycaption{Energy drained out of the SN core in the first 10 secs.
    The total energy (red solid line) is shown together with its
    visible (upper blue dashed line) and invisible (lower blue dashed
    line) components as a function of $\mmR$.}
\label{fig:losses}
\end{minipage}\hspace*{0.05\textwidth}
\begin{minipage}[t]{0.475\textwidth}
  \mycaption{Tau lepton number emission in the visible channel (blue
    dashed line) in the first 10 secs. The electron lepton number
    emission (green dot-dashed line) is also shown for comparison.}
\label{fig:lossesbis}
\end{minipage}
\end{figure}

Figures~\ref{fig:losses} and~\ref{fig:lossesbis} show the energy and
the tau lepton number drained out of the inner core in the first ten
seconds of evolution, separating the portion that is actually radiated
in the visible neutrino flux and the one that is irreparably lost into
the invisible channel provided by the bulk.  The first ten seconds are
the most interesting interval since this is the lapse of time during
which the relevant neutrino signal is produced.

As for the energy, one sees that a sizeable portion can be lost into
the invisible channel, especially for high values of $\mmR$.  At the
same time, the total amount of drained energy increases significantly
with $\mmR$, as a consequence of the faster cooling of the core.
However, balancing the two effects, what is important is that the
reduction of the portion emitted into the visible channel is always
limited to $\sim20\%$ of the standard case, which is well acceptable
in the light of the need of accounting for the \SN\ signal: no direct
upper bound on the parameter $\mmR$ then needs to be imposed.

\begin{figure}[!t]
\begin{center}
\epsfig{file=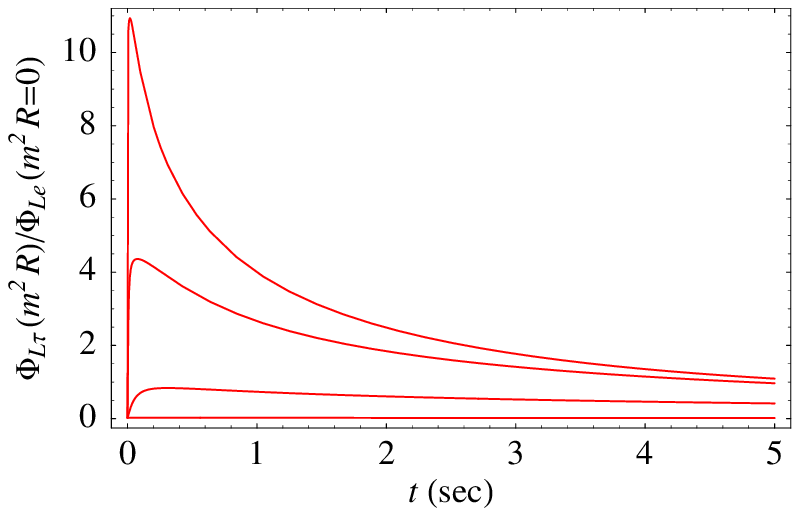,width=0.50\textwidth}
\end{center}
\mycaption{The tau lepton number flux versus time during the first 5
  secs, for increasing values of $\mmR=10^{-(12,~10,~8,~6)}$, from
  bottom to top. The normalization is given by the value of the
  electron lepton number flux in absence of new physics.}
\label{fig:lossvstime}
\end{figure}

The other significant result is the emission of a net tau lepton
number, counterbalancing the tau antineutrinos that escape into the
bulk.  The emission can grow up to values $\sim$ 10 times larger than
the corresponding $\nue$ flux, for $\mmR$ at the end of the considered
range.  As for its time distribution, \fig{lossvstime} shows that the
$\nut$ emission is concentrated in the very first seconds.  An
analogous behaviour is exhibited by the total energy emission, so that
we can foresee a peculiar time dependence of the final neutrino number
flux on Earth, more peaked at earlier times.  Letting aside this
general expectation, we focus in the following on time-integrated
quantities.

\medskip

Taking our outputs of the modified core evolution as the starting point, we now want to outline the subsequent vicissitudes and the ultimate fate of the neutrinos from the cooling phase in order to derive the possible consequences both for supernova physics and for the detectable signal on Earth.

First is the issue of how the energy flux and the sizeable $\nut$
net flux emitted from the inner core are redistributed by the
subsequent diffusion that takes place up to the neutrino sphere
(after which the neutrinos stream freely).  A detailed examination
of this phase would require studying the evolution of the mantle
also; see e.g. \cite{spectra}.  What we do is to assume that the
energy $E$ emitted from the inner core ends almost equiparted in
neutrinos and antineutrinos of all flavours, once the portion
carried by the lepton number excesses has been subtracted. On the
other hand, the (positive) lepton numbers $L_e$ and $L_\tau$ are
separately conserved in the reprocessing. With these ingredients
one can then determine the neutrino and antineutrino number fluxes
in each flavour: $n_{\bar\nu_j} \simeq \frac{1}{\langle
E_{\bar\nu_j}\rangle} \left(E-\sum_{i} \langle E_{\nu_i} \rangle
L_i\right)/ \left(3 + \sum_i\langle E_{\nu_i} \rangle/\langle
E_{\bar\nu_i} \rangle \right),\; n_{\nu_j} \simeq L_j+
n_{\bar\nu_j}  $, where $i,j=e,\mu,\tau$. For the average energies
of the different families we use the typical values $\langle
E_{\nu_e}\rangle \approx 13 \MeV$, $\langle E_{\nueb}\rangle
\approx 16 \MeV$ and $\langle
E_{\nu_{\mu,\tau},\bar\nu_{\mu,\tau}}\rangle \approx 23
\MeV$.\footnote{We are assuming that the average energies of the
  neutrinos at the neutrino sphere are not affected by the new physics
  effect, see~\cite{CCLR}.}
It has recently been pointed out \cite{KeilRaffelt} that the mean
energies of muon and tau neutrino and antineutrino can be closer
to the electron antineutrino's, at the level of less than $1.2\,
\langle E_{\nueb}\rangle $ during the core cooling phase.
Nevertheless, such a variation does not change our analysis.
 We will comment on this issue later on.

\begin{figure}
\begin{center}
\epsfig{file=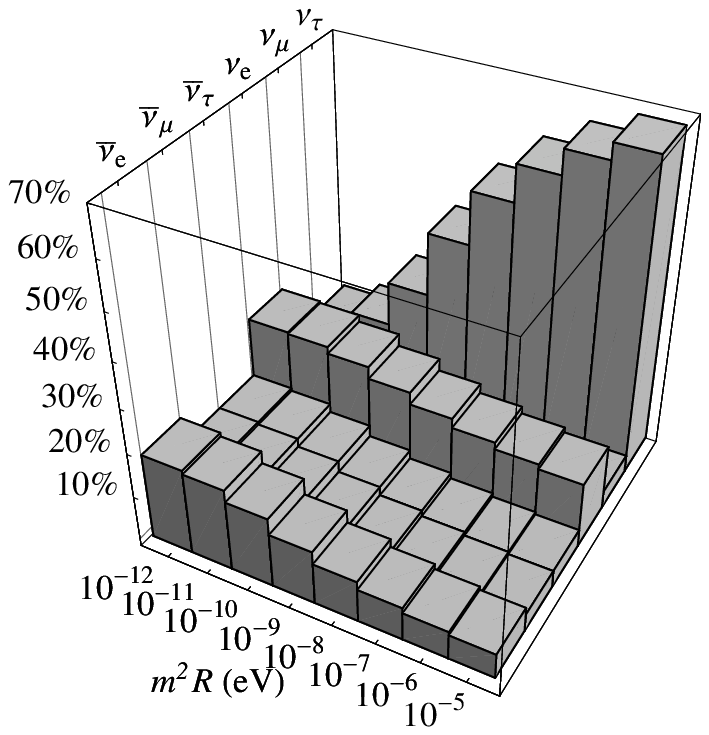,width=0.5\textwidth}
\end{center}
\mycaption{Indicative composition of the neutrino flux leaving the
  neutrino spheres as a function of $\mmR$.}
\label{fig:compplot1}
\end{figure}

Thus, coming out from the neutrino spheres, the flux displays a
peculiar composition (see \fig{compplot1}): for large values of
$\mmR$, the $\nut$'s are $\sim$ 5 times more than in the case without
new physics, at the expense of the $\nue$'s and of all flavours of
antineutrinos (all reduced by up to $\sim$ 60\%).  These reductions,
although not so large, go into the direction opposite to the need of
rejuvenating the stalling shock wave and thus helping the explosion,
since muon and tau neutrinos are much less efficient than the electron
ones in transferring their energy to the lingering matter.  On the
other hand, this qualitative argument does not imply a strong
drawback.  The actual explosion mechanism may be hidden in the
complexity of the system, whose simulation is still a non-trivial
task.

Next, neutrinos and antineutrinos undergo the matter flavour
oscillation in the peripheric low-density region of the star and the
vacuum flavour oscillations in the journey from the supernova to
Earth.  We illustrate the outcome in the case of normal neutrino mass
hierarchy. As for the oscillation parameters, we use $\theta_{23}
\simeq 45^\circ$~\cite{SK} and $\theta_{12} \simeq
32^\circ$~\cite{kamland}, while we consider values for
$\theta_{13}$ compatible with the Chooz bound~\cite{chooz}.  The first
transition at very high densities is produced by the small matter
potential difference between tau and muon neutrinos, which is due to
the mass difference of the corresponding leptons.  This transition is
found to be completely adiabatic and switches the
$\stackrel{\mbox{{\tiny (--)}}}{\nu}_{\mu}$, $\stackrel{\mbox{{\tiny
      (--)}}}{\nu}_{\tau}$ flavour eigenstates to the states rotated
by the angle $\theta_{23}$, labelled $\stackrel{\mbox{{\tiny
      (--)}}}{\nu}_{\mu'}$ and $\stackrel{\mbox{{\tiny
      (--)}}}{\nu}_{\tau'}$.  At lower densities, two more resonances
are met, due to the charged current contribution to the electron
matter potential. In the case of normal hierarchy, these resonances
are both active in the neutrino sector.  The $\nue$--$\nu_{\mu'}$
resonance turns out to be adiabatic under our assumptions while for
the $\nue$--$\nu_{\tau'}$ we consider a level crossing probability
$P_H$, which can vary from 0 to 1 depending on the value of
$\theta_{13}$.  For a more detailed analysis, we refer the reader
to~\cite{ALS}.  Finally, neglecting terms proportional to
$\sin{\theta_{13}}$, the neutrino and antineutrino fluxes $F$ reaching
the Earth's surface are related to the neutrino sphere fluxes $F^0$ by
the simple relations
\globallabel{eq:nufluxes}
\begin{align}
  F_e &= P_H \sud^2 F^0_e + \cud^2 F^0_{\mu} + (1-P_H) \sud^2
  F^0_{\tau} \mytag \\
  F_{\mu} &= \big( P_H \cud^2 \cdt^2 + (1-P_H) \sdt^2 \big) F^0_e +
  \sud^2 \cdt^2 F^0_{\mu}
  + \big( P_H \sdt^2 + (1-P_H) \cud^2 \cdt^2 \big) F^0_{\tau} \mytag\\
  F_{\tau} &= \big( P_H \cud^2 \sdt^2 + (1-P_H) \cdt^2 \big) F^0_e +
  \sud^2 \sdt^2 F^0_{\mu} + \big( P_H \cdt^2 + (1-P_H) \cud^2 \sdt^2
  \big) F^0_{\tau} \mytag
\end{align}
\globallabel{eq:antinufluxes}
\begin{align}
F_{\bar{e}} &= \cud^2 F^0_{\bar{e}} + \sud^2 F^0_{\bar{\tau}} \mytag\\
F_{\bar{\mu}} &= \sud^2 \cdt^2  F^0_{\bar{e}} + \sdt^2  F^0_{\bar{\mu}} + \cud^2 \cdt^2 F^0_{\bar{\tau}} \mytag \\
F_{\bar{\tau}} &= \sud^2 \sdt^2 F^0_{\bar{e}} + \cdt^2  F^0_{\bar{\mu}} + \cud^2 \sdt^2 F^0_{\bar{\tau}} \mytag
\end{align}
where $s_{ij}$ and $c_{ij}$ are the sine and cosine of the angle
$\theta_{ij}$.

\begin{figure}
\begin{center}
\epsfig{file=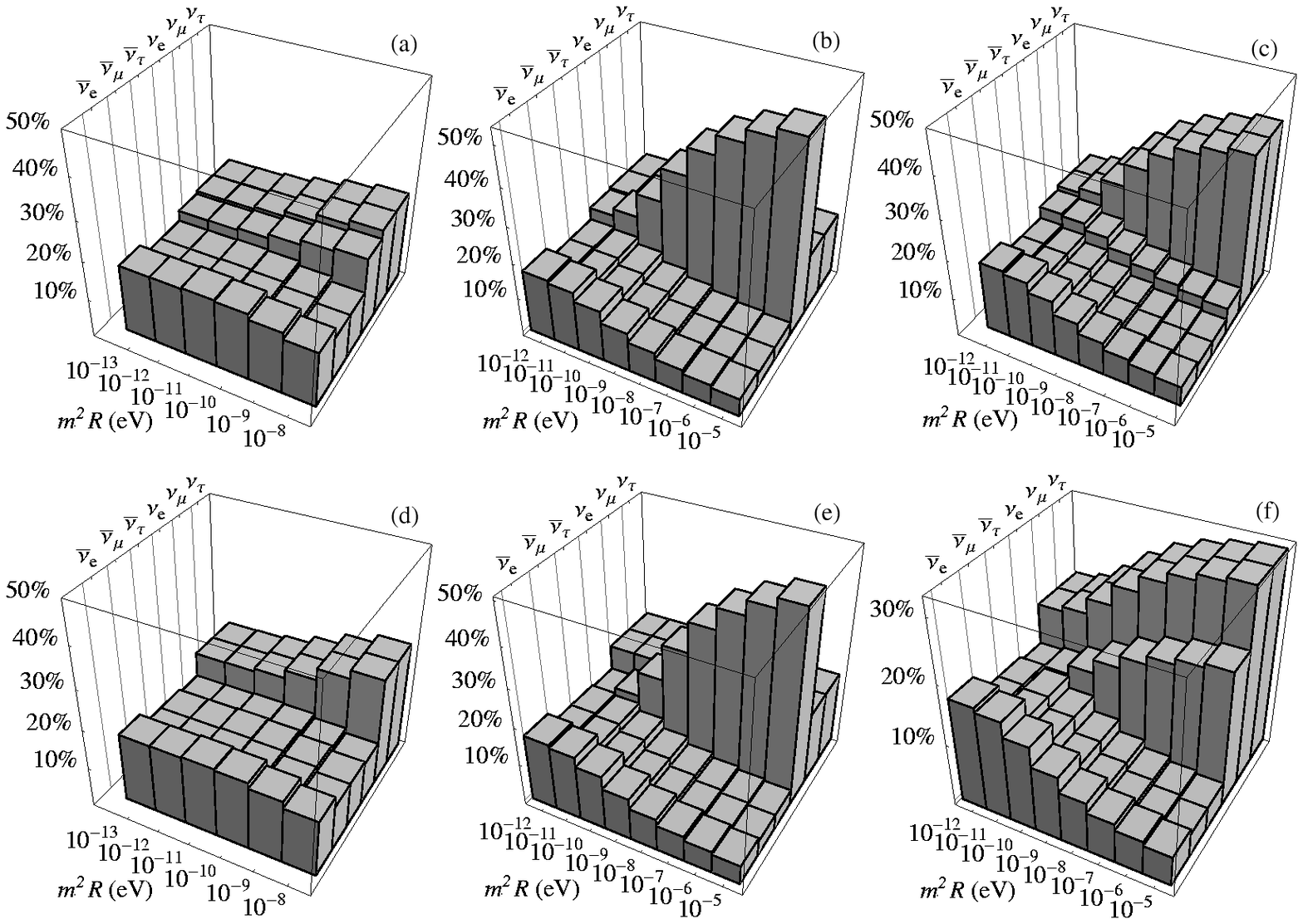,width=1\textwidth}
\end{center}
\mycaption{Indicative composition of the neutrino flux reaching the
  Earth's surface, as a function of $\mmR$, for the different cases of
  escape into the bulk: $\nue$ (first column), $\num$ (second column),
  $\nut$ (third column); for $P_H=1$ (first line) and for $P_H=0$
  (second line).}
\label{fig:signaltot}
\end{figure}

The indicative final composition of the flux reaching the surface
of the Earth is illustrated in \figs{signaltot}c and
\ref{fig:signaltot}f for the boundary values $P_H =1$ and $P_H
=0$, corresponding respectively to $\sin^2\theta_{13} \lesssim
10^{-5}$ and $\sin^2\theta_{13} \gtrsim 10^{-3}$.  In the first
case, the strong $\nut$ excess is distributed between muon and tau
neutrinos, while the electron fraction decreases.  The situation
is different in the adiabatic ($P_H=0$) case, where we expect an
higher $\nue$ component. A general important feature is the
$\nueb$ flux reduction that can amount to up to $\sim60\%$. Given
the limited statistics of the 1987 event, the overall
uncertainties on the expected neutrino fluxes (as predicted by
full simulations, also based on assumptions on the progenitor
star) and finally the approximate nature of the supernova core
evolution adopted here, such a reduction is still compatible with
present observations. Nevertheless, this feature is an interesting
and potentially challenging one, and deserves to be addressed more
closely in case of a future SN event with higher statistics. A
similar remark also holds for the time structure of the neutrino
signal mentioned above.

Our analysis has been limited so far to the case of tau neutrinos, but
can be readily extended to the case of mixing of muon neutrinos with
the bulk fermions.  Indeed, no relevant modifications are needed to
the model of evolution described in sec.~\ref{sec:model}, so that the
general conclusions reached for tau neutrinos are still valid.  In
particular, the indicative fluxes coming out of the neutrino spheres
are given in first approximation by a plot similar to the one in
\fig{compplot1}, with the excess in the $\num$ channel.  In turn,
using \eqs{nufluxes} and \eqs{antinufluxes}, one can determine the
compositions reaching the Earth's surface shown in \figs{signaltot}b
and~\ref{fig:signaltot}e; the value of $P_H$ is only marginally
important in this case.

Finally, we can recall the results of~\cite{CCLR} for the case of the
electron neutrinos oscillating into the bulk and derive the
compositions in \figs{signaltot}a and~\ref{fig:signaltot}d following
the same steps.

\medskip

Based on the above results, one can easily point out the most
important observable modifications of the expected flux in a detector.
First of all, as the net effect is always an increase in the total
emitted lepton number, a crucial signature is the enhanced ratio
between neutrinos and antineutrinos.  This is a general feature of
supernova neutrino oscillations into extra dimensions that does not
depend on which flavour has a significant mixing with bulk neutrinos.
What is case-dependent is which channel the enhancement shows up in.
First one can focus on the electron channel, which will have high
statistics in the future supernova events. Indeed, $\nueb$ are the
dominant signal in the present and in most of the future detectors,
while $\nue$ can be efficiently collected by large \v{C}erenkov
detectors, already existing (SK, SNO) or proposed (UNO), and
liquid-argon detectors (Icarus, LANNDD) \cite{cei}.  For several cases
in \fig{signaltot} the $\nue$/$\nueb$ ratio is significantly enlarged.
However, it is also clear from \fig{signaltot} that, depending on the
value of $P_H$ (i.e. of $\theta_{13}$), the effect of $\nue$ or $\nut$
conversion into the bulk could turn out to be irrelevant for the
$\nue$/$\nueb$ channel in some cases, so that it would be necessary to
measure the $\num$ or $\nut$ fluxes. This is a harder task, which is
however feasible at SNO, future detectors (UNO, LANNDD,
OMNIS)~\cite{cei} or even at large scintillator detectors (KamLAND,
Borexino~\cite{beacom}).  On the other hand, once an anomalous
neutrino/antineutrino ratio is measured, how is it possible to
distinguish which flavour does mix with the bulk?  The answer lies in
the energy spectrum.  In fact, since the electron neutrinos are less
energetic than the muon and tau ones when they leave the neutrino
spheres, the effect of the $\nue \leftrightarrow \nu_{\mu,\tau}$
oscillation is to harden the $\nue$ spectrum on the Earth.  The
conversion into the bulk of the electron flavour, increasing the
$\nue$ component flowing in the SN mantle, reduces this hardness,
while the muon or tau conversion enhances it.

We have illustrated our results for a specific point in the parameter
space of the standard neutrino oscillations. The residual dependence
on $\theta_{12}$ (and $\theta_{23}$) is mild. On the other hand, the
distribution of the neutrino flux enhancement in the three flavours
depends significantly on the neutrino mass pattern. Disentangling the
flavour structure of the mixing with bulk neutrinos would therefore
require the knowledge of sign($\Delta m^2_{23}$)~\cite{nufact}.

As already mentioned, another source of uncertainty is the precise
value of the mean energies of the neutrinos coming out from their
neutrino spheres. We checked that the adoption of a $\langle
E_{\nu_{\mu,\tau},\bar\nu_{\mu,\tau}}\rangle$ closer to $\langle
E_{\bar\nu_{e}}\rangle$, as suggested in~\cite{KeilRaffelt},
yields only minor modifications of our result. Namely, each column
in \fig{compplot1} and \figs{signaltot} is affected at most by a
variation of $5 - 10 \%$ in composition, within the uncertainties
of our simplified model. Thus, the analysis is unchanged and the
general trend confirmed: we only remark that the closer the mean
energies, the smaller the differences among the channels, although
preserving the features described above.

Finally, in order to predict the actual signal in the detectors, one
should also consider matter-induced oscillations inside the Earth.
This effect has the power of dramatically changing the expected
signal, but of course depends on the geographical position of the
detectors at the time of arrival of the signal. For a given path in
the Earth, it is not difficult to include the matter effect (see
\cite{flavour,ALS}).

\medskip

While we have so far considered the case in which the neutrino escape
is dominated by the oscillations in the bulk of one of the three
neutrinos, it is also possible that all SM neutrinos are involved in
the energy loss.  In this case, the neutrino flux on the Earth would
still be enhanced over the antineutrino one. However, the neutrino
flux composition would also be determined by the relative size of the
three mixing parameters $m_e$, $m_\mu$, $m_\tau$. The evolutions of
the three neutrino abundances in the core would be coupled through the
temperature and the matter potential.  The feedback mechanisms would
however still take place, but we expect the $m_i$--$R$ parameter space
for the three channels to be slightly reduced.

\section{Conclusions}
\label{sec:conclusions}

We dealt with the mixing of the Standard Model neutrinos with bulk
sterile fermion fields, a natural and general feature of models with
extra dimensions.  In particular, we focused on the muon and tau case,
while the electron case was addressed in~\cite{CCLR}.  This mixing
provides an unconventional escape channel for neutrinos in the
supernova core during the cooling phase, which could in principle give
strong bounds on the parameters of the extra dimensions when facing
the observed \SN\ signal.  In order to simplify the core dynamics, we
restricted the analysis to the case of $\nut$, but we easily completed
it to the case of $\num$.

Given the sign of the matter potential, the process consists in the
escape of tau antineutrinos into the bulk, acting in parallel with
diffusion.  We showed that a feedback mechanism turns on, preventing
extreme and unacceptable modifications of the evolution to occur (in
particular, the characteristic time scale of deleptonization and
cooling, of order 10 sec, is preserved).  The direct bound on the
parameters of extra dimensions can then be relaxed by about 4 orders
of magnitude.  The upper limit on the relevant combination $\mmR$ is
lifted from $10^{-10} \eV$ to $5\times 10^{-5} \eV$, the new limit now
coming from subleading effects.  In \fig{bounds1}, we showed the
allowed region in the $m$--$R$ plane.  The feedback is mainly due to
the building up of a degenerate sea of trapped $\nut$ and thus of a
positive tau neutrino chemical potential that suppresses the number of
$\nutb$'s candidate to escape.  For comparison, in the $\nue$ (and
$\nueb$) case in~\cite{CCLR} this kind of feedback acts together with
the feedback on the matter potential in a sizeable portion of the
core.

Although in such a `safe' way, the supernova core evolution is
significantly affected by the escape process, and we studied the main
modifications.  First of all, the cooling is sped up with respect to
the standard case.  Secondly, a sizeable net tau lepton number is
emitted by the core, concentrated in the very first seconds.
Therefore a neutrino number flux more peaked at earlier times can be
predicted.

Next we speculated on the phenomenological consequences on the revival
of the shock and on the neutrino signal reaching the Earth.  As for
the first point, since the $\nue$ flux is lowered by the reshuffling
of the neutrino fluxes (although not largely), we suggested that the
$\num$, $\nut$ escape may reduce the energy transfer to matter and
thus the push to the explosion.  As for the second point, the
compatibility with the \SN\ signal is preserved once one is willing to
accept a certain reduction of the $\nueb$ component.  Collecting the
results for the $\nue$, $\num$ and $\nut$ escape cases, we indicated
as signatures, in cases of future galactic SN events, a general
dominance of neutrino flux over antineutrinos and, depending on the
flavour that mixes with bulk fields, some peculiar structures of the
relative enhancements in the neutrino fluxes and a harder or softer
$\nue$ spectrum.

\section*{Acknowledgments}

This work has been partially supported by MIUR and by the EU under TMR
contract HPRN--CT--2000--00148.  Part of the work of A.R. was done
while at Scuola Normale Superiore.  M.C. acknowledges the hospitality
of the CERN Theory Division, where this work was completed.

\end{document}